\documentstyle[preprint,tighten,prc,epsf,eqsecnum,aps]{revtex}

\begin{document}
\title{ Projectile fragmentation of $^{129}$Xe at E$_{lab}$=790 A$\cdot$MeV}
\author {J.~Reinhold\thanks{Present address: Physics Division,
         Argonne National Laboratory, 9700 South Cass Avenue, IL 60439, USA}, 
         J.~Friese, H.-J.~K\"orner, R.~Schneider,
         and K.~Zeitelhack}
\address{Physikdepartment E12, Technische Universit\"at M\"unchen, 
         D-85747 Garching, Germany}
\author{H.~Geissel, A.~Magel, G.~M\"unzenberg, and K.~S\"ummerer}
\address{Gesellschaft f\"ur Schwerionenforschung, D-64220 Darmstadt, Germany}

\date{\today}
\maketitle

\begin{abstract} We have measured production yields and longitudinal
momentum distributions of projectile-like fragments in the reaction
$^{129}$Xe + $^{27}$Al at an energy of E$_{lab}$=790 A$\cdot$MeV.
Production cross sections higher than expected from systematics
were observed for nuclei in the neutron-deficient tails of the isotopic
distributions. A comparison with previously measured data from the
fragmentation of $^{136}$Xe ions shows that the production yields strongly
depend on the neutron excess of the projectile with respect to the line of
$\beta$-stability. The momentum distributions exhibit a dependence on the
fragment neutron-to-proton ratio in isobaric chains, which was not
expected from systematics so far. This can be interpreted by a higher
excitation of the projectile during the formation of neutron-deficient
fragments.\pacs{} \end{abstract} 
\narrowtext
\section{Introduction}
Numerous studies have shown that projectile fragmentation at high energies
is a powerful tool to produce intensive beams of exotic 
nuclei even close to the driplines~\cite{geissel95}.
In order to get a quantitative estimate of
the production cross sections of exotic nuclei, empirical parameterizations
have been developped and fitted to the available experimental data
(e.g.~\cite{rudstam66,EPAX,Sihver_epax}). Important for the planning of
future experiments is, whether these parameterizations yield reliable
predictions even for very exotic nuclei in the tails of the isotopic
distributions, where experimental data are sparse. Also of considerable
interest for the study of nuclei at the borderline of stability is whether
the isotopic yields may be influenced by the use of appropriate
neutron-rich or neutron-deficient projectiles. Experiments which measured 
the fragmentation of
different
isobaric projectiles (or targets) showed a shift of the fragment distributions,
which is
related to the projectile (target) neutron or proton excess
relative to the line of
$\beta$-stability~\cite{48Ca_212AMeV,memory_porile,ku_karol77}. This
"memory effect" has been
included quantitatively in the EPAX parameterization~\cite{EPAX},
but only few data from light projectiles were available to fit the
corresponding parameters. Nevertheless this parameterization described
successfully the fragmentation yields from the neutron rich 
isotope $^{86}$Kr~\cite{weber92}. More recently, Pfaff et al. came to very 
similar conclusions from their studies of $^{78,86}$Kr fragmentation at 
intermediate energies~\cite{pfaff96}. 

In an earlier experiment we have measured the projectile
fragmentation of $^{136}$Xe~\cite{136Xe},
the most neutron-rich stable xenon isotope. In
the present experiment the fragmentation of $^{129}$Xe was investigated
with several questions to be addressed. Firstly the memory effect in the mass
region A$>$100 should be verified by comparison to the $^{136}$Xe
data. Furthermore, the very neutron-deficient tails of the fragment
distributions were to be studied to investigate the production of nuclei near
the proton dripline in the mass region A$\approx$100.

Apart from the study of isotopic distributions, a further objective was to
measure the momentum distributions of the fragments with high precision. This
should allow to get more insight into the underlying reaction mechanism,
which is commonly described as a two-step process~\cite{serber47}: In a
first collision phase nucleons are abraded from the projectile by
individual nucleon-nucleon scattering processes. An excited prefragment is
left, which then deexcites by the emission of nucleons and $\gamma$-rays
in an evaporation-like cascade process to form the fragment finally 
observed.
Since models which ignore the
specific characteristics of the nucleon-nucleon scattering processes in the
first reaction step, the so-called abrasion-ablation models
(e.g.~\cite{abrabla} and references therein), are quite successful in
describing the isotopic distributions, further observables have to be
measured to distinguish between different models.
The deexcitation of the prefragments should
be governed by an isotropic emission of
particles and thus only influence the width of the momentum distributions. 
In contrast, the mean value of the fragment
momentum should be sensitive only to the collision phase.
Therefore the
clue to disentangle the different reaction steps and their influence on the
final fragment formation is a detailed study of the momentum distributions.

In this paper first the experimental procedure will be presented. Then the
results will be discussed on a phenomenological basis. A more detailed
discussion with a comparison to an intranuclear-cascade model will be given
in a forthcoming paper. 
\begin{figure}
\begin{center}
\leavevmode
\epsfxsize=10cm
\epsffile{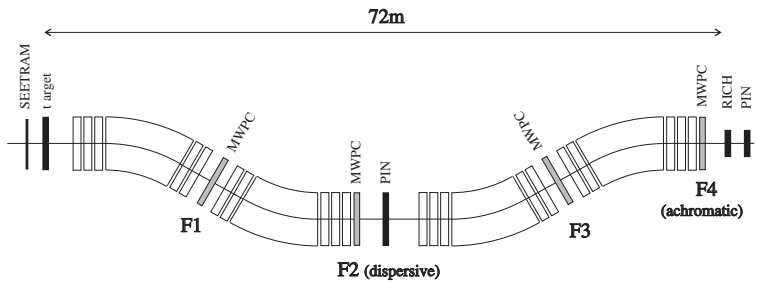}
\end{center}
\caption{Experimental setup at the fragment separator FRS.  The
multi-wire proportional counters (MWPC) at focal points F1 to F4 were
only used to tune the separator and were removed from the beam during
the actual measurements. The silicon detectors (PIN) and the
\v{C}erenkov detector (RICH) were positioned in the
momentum-dispersive focus F2 and in the achromatic focus F4,
respectively.}
\label{fig:setup}
\end{figure}
\section{Experimental procedure}
The experiment was performed at the GSI facility at Darmstadt, Germany.
$^{129}$Xe ions were accelerated in the heavy-ion synchrotron SIS to an
energy of 790 A$\cdot$MeV. Spills with intensities of up to 10$^8$ ions
lasting for about 3-4 seconds with a repetition rate of $1/5$ s$^{-1}$ were
focussed onto an (803$\pm$2) mg/cm$^2$ aluminum target in front of the
fragment separator FRS~\cite{FRS}. The experimental setup is sketched in
Fig.~\ref{fig:setup}. 
The primary beam intensity was determined from the
current induced by delta electrons in an aluminum converter
foil (SEETRAM~\cite{SEETRAM}) in front of the target.
This detector had an areal thickness of
9~mg/cm$^2$ aluminum, thus increasing the total target thickness to
812~mg/cm$^2$ aluminum. With a total reaction cross section of
$\sigma_{total}=3.5$~b calculated from an empirical
parameterization~\cite{kox} the
total reaction rate of the primary beam in this target was 6.3\%. Several
large-area multi-wire proportional counters (MWPC) were available at the
different focal planes. These could be moved into the beam to tune the
separator. The projectile-like fragments produced were separated from the
beam in the first half of the FRS. In the momentum-dispersive central focal
plane (F2) their positions and thus their magnetic rigidities were measured
with a segmented silicon detector array. This detector consisted of 
sixty-four
220$\mu$m thick silicon photodiodes with an active area of $10\cdot 10$~mm$^2$.
They were arranged into four subsequent layers of $16\cdot 1$~cm$^2$,
which were shifted against each preceeding layer by 2.5~mm in horizontal
(bending) direction. This results in a position resolution of $\pm 1.25$~mm,
which translates into a momentum resolution of $\Delta p/p = 1.7\cdot
10^{-4}$. This detector system also served to determine the nuclear charge
by measuring the energy-loss of the fragments. The second half of the separator
was tuned such as to produce an achromatic focus in the final focal plane
F4, which means that at this focus the fragment positions are independent
of their momenta. Here their velocities were measured with a
ring-imaging \v{C}erenkov detector~\cite{RICH}. A second silicon detector array
behind the \v{C}erenkov detector with an active area of $60\cdot 30$~mm$^2$ 
served as an additional trigger detector.

The SEETRAM beam intensity monitor was calibrated at low beam currents ($\leq
2\cdot 10^5$ ions/s) by counting individual beam particles
with a scintillator which could be moved into the
beam. We estimate the error of the beam intensity to be 5--17\%, resulting 
from both the calibration error and the counting statistics of the SEETRAM 
current digitizer.

The magnetic dipole fields were measured by Hall probes with an accuracy of
$\Delta$B=10$^{-4}$~T. The dispersion in the central focal plane F2 was
determined to be $D=\Delta x /(\Delta B\rho /B\rho_0)=(74.58\pm 0.65)$mm/\%
by measuring the position of the primary beam for several field settings.
With a beam spot of $\Delta x\approx\pm 2.7$~mm at the target and a position
resolution of the silicon detector at the central focal plane F2 of $\Delta
x=\pm 1.25$~mm this allowed to determine the momentum for fragments with
known ionic charge with an accuracy of $\Delta |\vec P| \approx$60~MeV/c. 

\begin{figure}
\begin{center}
\leavevmode
\epsfxsize=10cm
\epsffile{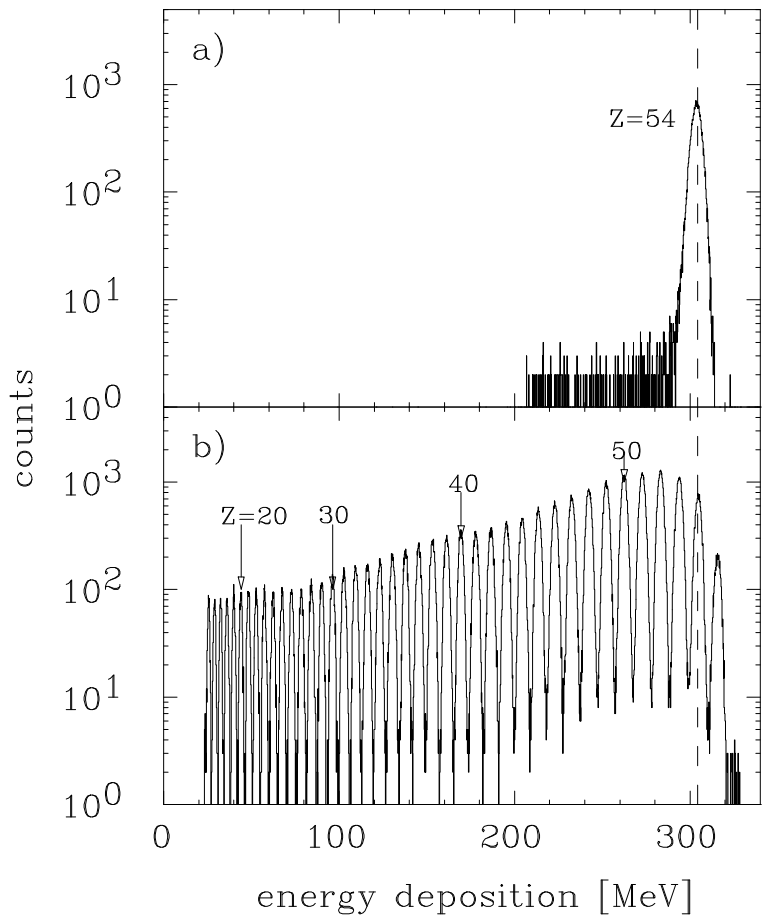}
\end{center}
\caption{Response of the silicon detector array at F2: {\bf
a)}: the primary beam with Z=54 was used for calibration.
Shown is the response of a single detector chip in one
layer. The low-energy tail results from fragments. {\bf b)}:
charge resolution of the entire detector system for
fragments with B$\rho = (10.486\pm 0.105)$Tm, corresponding
to $A/Q\approx (2.24\pm 0.02)$.} 
\label{fig:zeich}
\end{figure}

The fragment nuclear charge number Z was determined from a fourfold 
energy-loss measurement in the silicon detector array at F2. This detector was
calibrated with the primary beam for the nuclear charge number Z=54
(Fig.~\ref{fig:zeich}a). At this energy 99\% of the ions are fully ionized
(Q=Ze)~\cite{chargestate}. Thus this measurement also determines the
ionic charge of the fragments. 

\begin{figure}
\begin{center}
\leavevmode
\epsfxsize=10cm
\epsffile{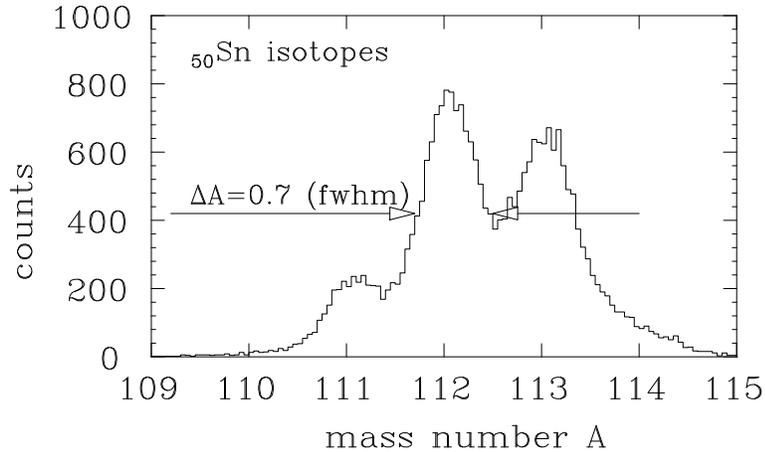}
\end{center}
\caption{Mass spectrum for tin isotopes with B$\rho =
(10.486\pm 0.105)$Tm.} 
\label{fig:masse}
\end{figure}
The velocity resolution achieved with the \v{C}erenkov detektor was
$\Delta\beta/\beta = 1.8\cdot 10^{-3}$~\cite{RICH}. Together with the
magnetic-rigidity and energy-loss measurements, the fragments could be
identified with respect to nuclear charge and mass number. 

As the spectrometer acceptance is about $\pm$1\% in momentum, several field
settings were necessary to cover the neutron-deficient area of the nuclear
chart in the projectile mass region up to the vicinity of the proton
dripline. In addition in one setting we selected fragments that were more
neutron-rich than
the projectile for comparison with an earlier experiment using a $^{136}$Xe
beam~\cite{136Xe}. The different field settings were each optimized for the
transmission of a specific tin isotope. 
\section{Results}
\subsection{Isotope Identification}

In Fig.~\ref{fig:zeich} an energy-loss spectrum obtained with the silicon
detector at F2 is shown. The upper part shows the response of a single
diode to the primary beam with Z=54. The low-energy
tail in this spectrum results from fragmentation products
with magnetic rigidities
similar to that of the beam.
The four layers of the Si
detector allow a fourfold
energy-loss measurement, which further improves the resolution.

The lower
part of Fig.~\ref{fig:zeich}
shows the response of the entire detector system to fragments measured
with the spectrometer setting optimized for the transmission of
the isotope $^{112}$Sn. 
The different nuclear charges are clearly resolved. As the fragment
velocities are close to the beam velocity, which again is near the 
minimum-ionizing region,
the absolute value of the nuclear charge number can be
obtained by direct comparison with the beam spectrum. Applying three-sigma
window conditions on this charge spectrum in the further analysis, the
probability for misidentification in charge number is below 2\%. Neglecting
this small ambiguity, integer numbers were assigned to the individual
fragments in the subsequent analysis. 

For fragments with nuclear charge numbers below Z=40 the different
energy-losses, compared to the nominal fragment,
lead to a horizontal displacement of their foci at
the final 
focal plane F4 which is too large to be accepted by the \v{C}erenkov
detector. Therefore only fragments with Z$\geq$40 could be identified. The
response of the \v{C}erenkov detector and its analysis is discussed
elsewhere~\cite{RICH}. A resulting mass spectrum obtained by combining the
charge, velocity and position measurements is shown in Fig.~\ref{fig:masse}
for tin isotopes. The appearance of mainly three isotopes reflects the
spectrometer momentum acceptance of $\pm$1\% which corresponds to $\pm$1
mass unit for isotopes with nuclear mass number A$\approx$100. The relative
intensities of the three tin isotopes in Fig.~\ref{fig:masse} are dominated
by their different transmissions through the spectrometer rather than their
production cross sections. The upper part of Fig.~\ref{fig:beta_ort} shows
the measured velocity vs. the position at F2 for the same isotopic
distribution. This demonstrates that the width of the momentum
distributions is of the same order as the spectrometer acceptance and
therefore only the central isotope (here $^{112}$Sn) is expected to have a
transmission near 100\%. 

\subsection{Determination~of~Cross~Sections}
Individual isotopic cross sections were determined from the number of
counts in the projected position spectra like the ones shown in
the bottom part of Fig.~\ref{fig:beta_ort}. This
allowed also to determine the ion-optical transmission. The
transmission was derived from the ratio of the measured counts to the area 
of a fitted
Gaussian function folded with a rectangular distribution. The
rectangular distribution accounts for the different energy losses of
projectile and fragment in the target. Its width was fixed and determined
from energy loss calculations~\cite{ATIMA}. As there is no comparable
information perpendicular to the bending direction, transmission
losses in vertical direction where determined by occasionally starting the
data acquisition with the larger MWPC's. The ratio of
correlated to uncorrelated events between these detectors and the silicon
detector indicated an additional transmission loss of 5\%.

The overall efficiency, including ionic charge changes, secondary reactions
in the detectors, and deadtime was approximately 60\%. The
experimental errors are dominated by the beam intensity
monitor (5--17\%) and the
transmission determination (10--50\%). For isotopes with low
transmission no reliable fit of the position distributions was possible
(see e.g. the $^{111}$Sn distribution in Fig.~\ref{fig:beta_ort}).
But in most cases
such isotopes were observed with higher transmission in an adjacent field
setting. For isotopes where the distributions could be fitted in two 
settings, the results agree within the extracted errors. Only for 
distributions with low statistics, which did not allow a reliable fit, a 
Monte Carlo simulation~\cite{mocadi} of the spectrometer transmission
was used. This 
simulation agreed with the measured transmissions within about 5\%.

Cross sections down to 1nb could be determined in the mass region
$80<A<129$ with nuclear charge numbers Z from 40 to 55. All derived cross
sections are given in Table~\ref{tbl:results}. The charge-pickup process
leading to fragments with Z$_{frag}$=55 has been discussed in a separate
paper~\cite{CEX}. 
\subsection{Momentum~Distributions}
\begin{figure}
\begin{center}
\leavevmode
\epsfxsize=7cm
\epsffile{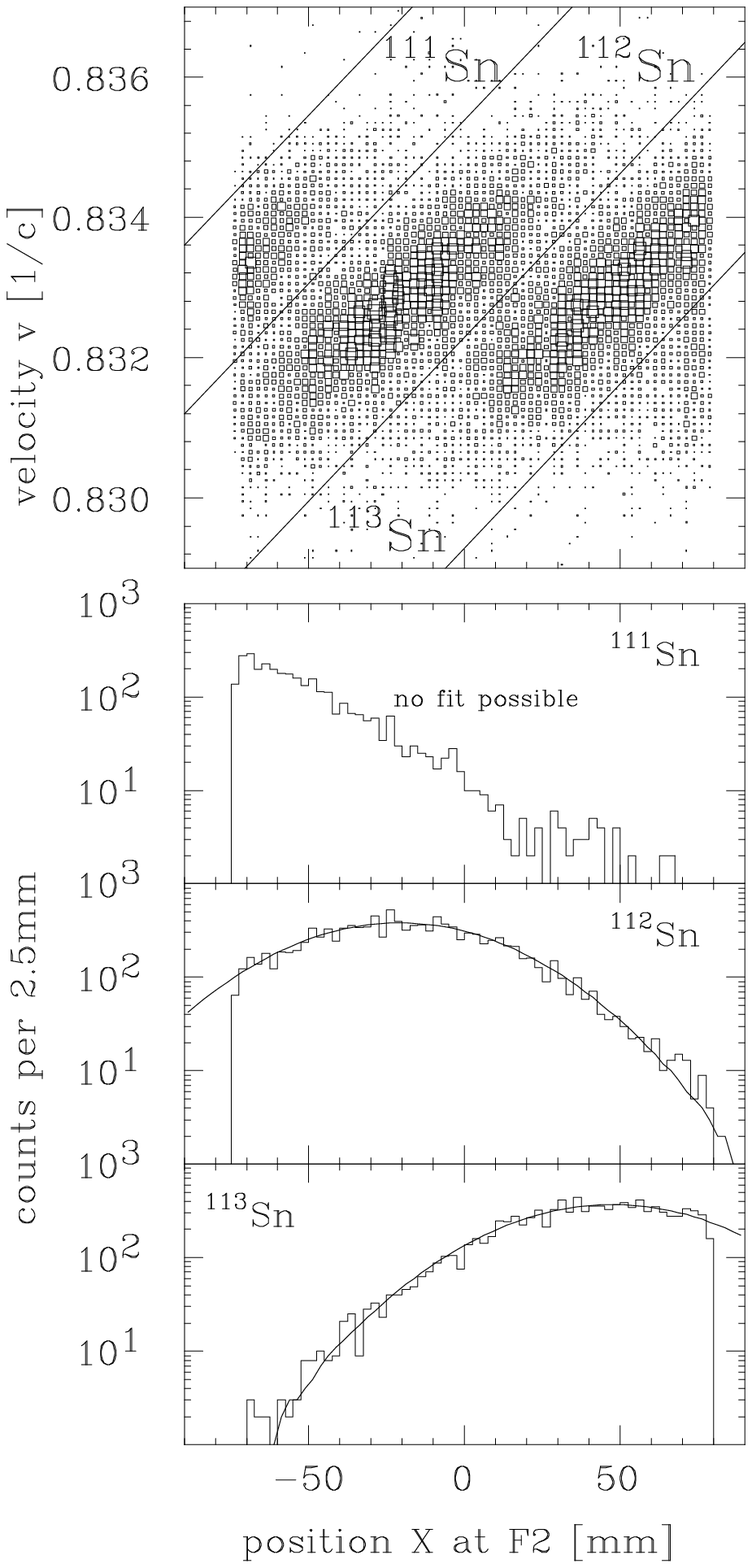}
\end{center}
\caption{Top: Measured velocity vs. position at the
dispersive focus F2 for tin isotopes with B$\rho=(10.486\pm
0.105)$Tm.\protect\\ Bottom: Projections on the horizontal
axis for the three isotopes shown in the top figure. The
spectra were fitted by a Gaussian folded with a rectangular
distribution. For the isotope $^{111}$Sn no reliable fit was
possible, but it could be measured with an adjacent field
setting.} 
\label{fig:beta_ort}
\end{figure}

The average momentum and the width of the fragment momentum distributions
were determined from the position spectra (Fig.~\ref{fig:beta_ort})
assuming a Gaussian momentum distribution 
\noindent folded with a rectangular
distribution as described in section 3.2. Both quantities were transformed
into the projectile restframe.
The velocity change of projectile and fragment
due to energy loss in the target was
determined by energy loss calculations~\cite{ATIMA}, which  have been shown
to be accurate to about 2\%~\cite{FRS}.
Due to the large fragment momentum in the
laboratory frame the contribution of transverse momentum components to the total
momentum is smaller than the spectrometer resolution.
Thus the experiment was only
sensitive to the longitudinal (in beam direction) momentum change.
Transverse components should in principal be measurable by determining the
fragment angular distributions, but were not achievable with the present
layout of the silicon detector used in the central focal plane. 

The field measurement, the position measurement and the calibration of
the dispersion contribute to the error with $\Delta P/P = 1\cdot
10^{-4}, 4\cdot 10^{-4}$, and $8.8\cdot 10^{-5}$, respectively. Additional
contributions arise from uncertainties in the fitting procedure, i.e.
for distributions with low statistics or low transmission. The results
are summarized in Table~\ref{tbl:results}.
\begin{figure}
\begin{center}
\leavevmode
\epsfxsize=8cm
\epsffile{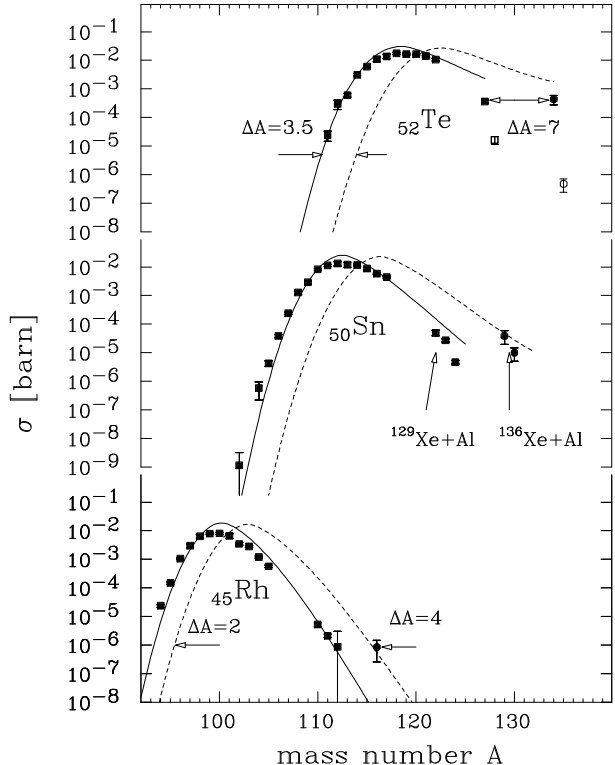}
\end{center}
\caption{Measured production cross sections for 760 A$\cdot$MeV
$^{136}$Xe+Al (~\protect\cite{136Xe}, circles) and 790 A$\cdot$MeV
$^{129}$Xe+Al (this work, squares) compared to the EPAX
parameterization~\protect\cite{EPAX}. Note that the
formation of the most neutron-rich tellurium isotopes
(marked with open symbols) requires charge exchange
reactions during fragment formation, as $\Delta$A=1 but
$\Delta$Z=2.} 
\label{fig:memory}
\end{figure}
\section{Discussion}
\subsection{Cross~Sections}
Representative examples of isotopic distributions for the elements of Te, 
Sn, and Rh
are shown in 
Fig.~\ref{fig:memory}. They exhibit the typical Gaussian like shapes,
where the slope of the neutron-rich tail is less steep than that of the
neutron-deficient side. In general these distributions are reproduced by
the empirical parameterization EPAX~\cite{EPAX} (full line) with respect to
the
position of their maxima and their shape. The most significant deviations
occur for neutron-rich fragments with masses close to that of the 
projectile.
The production of these fragments should be governed by the "cold"
removal of protons, because excitation of the projectile dominantly
leads to neutron evaporation, which is not hampered by the Coulomb
barrier. Therefore a parameterization, which mainly has been fitted to
spallation data, is not expected to describe these specific reaction
channels. This had already been observed in other
experiments~\cite{proton_removal}, including our previous
measurement with the $^{136}$Xe projectile~\cite{136Xe}
(the latter is shown
in Fig.~\ref{fig:memory} with the circled symbols and dashed lines).
Aside from this deficiency the shift of the isotopic distributions is
described satisfactorily. However, a direct comparison of the two projectiles
is only possible for
the neutron-rich tail due to experimental constraints in the $^{136}$Xe
experiment. As indicated by the arrows in Fig.~\ref{fig:memory},
the excess of seven neutrons for the $^{136}$Xe projectile
is fully preserved for fragments close to the projectile (tellurium
isotopes in Fig.~\ref{fig:memory}), and even for fragments that have
lost as much as 20 nucleons a shift of the neutron-rich tail of the
isotopic distribution by four mass units is clearly visible (rhodium
isotopes in Fig.~\ref{fig:memory}). Up to now this memory effect has only 
been observed for lighter 
nuclei~\cite{48Ca_212AMeV,memory_porile,ku_karol77}.
Therefore our data corroborate the quantitative description of the "memory 
effect" contained in EPAX
for  heavier projectiles. 

Particular attention should be paid to
the neutron-deficient tin isotopes,
where due to the experimental procedure (the separator was always
optimized for 
the transmission of a tin isotope) the lowest cross sections could be
measured. The slope of the distribution seems to differ significantly
from the EPAX parameterization, a fact that has also been observed for
neutron-deficient isotopes produced in the fragmentation of $^{58}$Ni
projectiles~\cite{blank94a}. These observations, and the memory effect,
which is also
predicted for neutron-deficient projectiles, opens up the prospect to 
produce
the doubly magic nucleus
$^{100}$Sn
by fragmentation  of $^{124}$Xe, which is the most
neutron-deficient xenon isotope available as a projectile. In the meantime
this experiment
was performed successfully and results have been presented
in Refs.~\cite{100Sn,SELMA94,Niigata94}.
\subsection{Longitudinal~Momentum~Distributions}
\begin{figure}
\begin{center}
\leavevmode
\epsfxsize=10cm
\epsffile{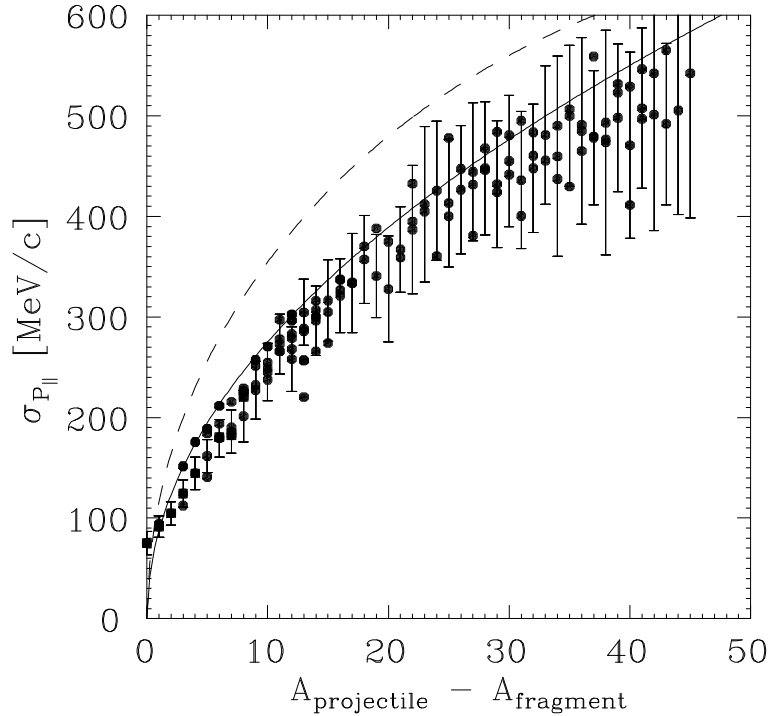}
\end{center}
\caption{Width parameters of the longitudinal momentum
distributions.
For clarity only one typical error bar is shown for each 
mass. For individual errors see Tab.~\ref{tbl:results}.
Full line: Empirical parameterization by
Morrissey~\protect\cite{morrissey}. Dashed line: Goldhaber
model~\protect\cite{goldhaber} (see text).} 
\label{fig:goldhaber}
\end{figure}

Fig.~\ref{fig:goldhaber} shows the measured widths of the fragment momentum
distributions in the projectile restframe.
In previous studies, authors have compared measured momentum widths (see 
e.g.~\cite{christie93,greiner75}) to the predictions of the Goldhaber 
model~\cite{goldhaber}. This "sudden break-up" model predicts the momentum 
width $\sigma_{\Vert}$ of a break-up residue (a {\em prefragment} in our 
terminology, see Sect. 1) with mass A$_{pf}$ to obey the equation
\begin{equation}\label{eqn:goldhaber}
\sigma_{P_{\Vert}} = \sigma_0\cdot
\sqrt{\frac{A_{pf}(A_{proj}-A_{pf})}{A_{proj}-1} }
\qquad\mbox{ ,}
\end{equation}
where $\sigma_0=\sqrt{1/5} P_{Fermi}$.
Here $P_{\Vert}$ is the longitudinal momentum of the fragments in the projectile
restframe, $P_{Fermi}$ the Fermi momentum of nucleons in the projectile,
and $A_{proj}$ and $A_{pf}$ the mass number of the projectile and
prefragment, respectively. A numerical value of 
$P_{Fermi}$=260MeV/c, can be taken from
quasielastic electron scattering data~\cite{moniz71}. 

Since in our experiment we mainly observe fragments which are produced by 
evaporation cascades from the prefragments, we cannot expect their momentum 
widths to follow Eq.~(\ref{eqn:goldhaber}). Consequently, the Goldhaber 
prediction (dashed line in Fig. 6) clearly disagrees with the data. Only in 
cases where a surviving prefragment can be observed (e.g. proton-
\noindent  removal 
channels in the fragmentation of neutron-rich projectiles, 
Refs.~\cite{proton_removal,hanelt93}) good agreement with the Goldhaber 
model is observed.

The full line in Fig.~\ref{fig:goldhaber} shows an empirical parameterization,
\begin{equation}\label{eqn:morrissey_gold}
\sigma_{P_{\Vert}}\approx 87MeV/c\cdot\sqrt{A_{proj}-A_{frag}} 
\qquad\mbox{ ,}
\end{equation}
which was derived from a large compilation
of available experimental data~\cite{morrissey} and gives a quite adequate 
representation of our measured data.

\begin{figure}
\begin{center}
\leavevmode
\epsfxsize=10cm
\epsffile{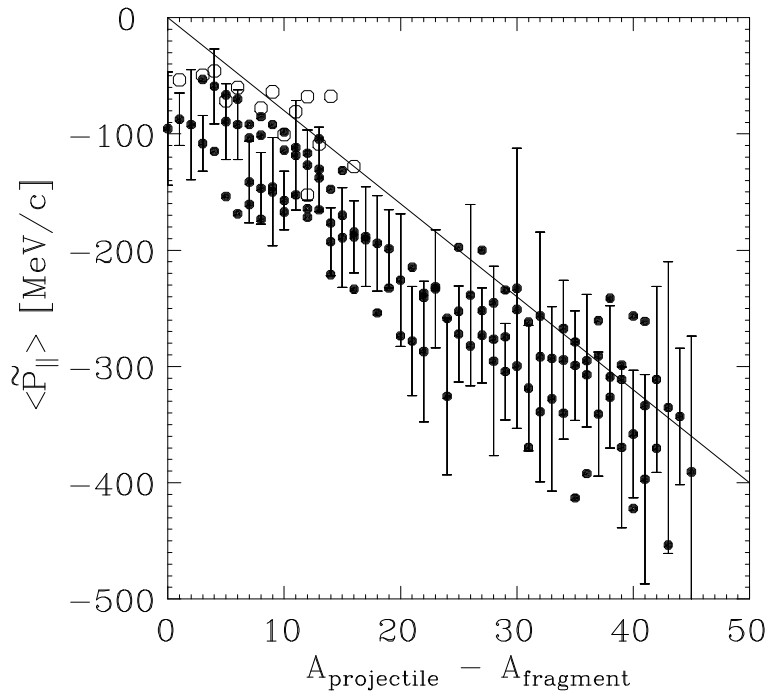}
\end{center}
\caption{Mean longitudinal ``momentum'' compared with the
systematic from Morrissey \protect\cite{morrissey} (Note
that $\left< \tilde P_{\Vert}\right>$ is not the real
fragment momentum but its mean velocity multiplied with a
constant factor. For details see text). Open circles denote
fragments measured in the most neutron rich setting.
For clarity only one typical error bar is shown for each 
mass. For individual errors see Tab.~\ref{tbl:results}.
} 
\label{fig:morrissey}
\end{figure}

The above mentioned compilation~\cite{morrissey} also gives a
parameterization of the mean longitudinal momentum $\langle\tilde
P_{\Vert}\rangle = \Delta A\cdot 8MeV/c$, where 
the definition
\begin{equation}\label{eqn:morrissey}
\left<\tilde P_{\Vert}\right> \equiv m_{proj}c  
\left<\beta_{\Vert}\right> \frac{\beta\gamma}{\gamma+1} \qquad\mbox{,}
\end{equation}
was used.
Here $\left<\beta_{\Vert}\right>$ is the average fragment velocity in
the projectile restframe, $\beta$ and $\gamma$ are the velocity of the 
projectile and its Lorentz factor, respectively.
Thus $\langle\tilde
P_{\Vert}\rangle$ represents the fragment velocity distribution rather than 
the momentum distribution.
It has been pointed out, that under certain 
assumptions $\langle\tilde
P_{\Vert}\rangle$ may be a measure of the excitation 
energy~\cite{morrissey} of the projectile or prefragment.
We use this
expression for the "average momentum" to compare our data to the 
systematics
in Fig.~\ref{fig:morrissey}.

In general the slope of the data is reproduced by
the parameterization (solid line in Fig.~\ref{fig:morrissey}), however there
are significant deviations for the individual data points.
The best
\noindent agreement is 
obtained for
the most neutron-rich fragments measured, indicated by open symbols in
Fig.~\ref{fig:morrissey}. 
The clue for an explanation of this behaviour is the origin of 
the parameterization. 
Most of the data referred to in~\cite{morrissey} were obtained in target 
fragmentation experiments. Thus only isotopes with lifetimes sufficiently 
long to be detected with radiochemical methods could be identified. This 
restricts the accessible area of the nuclear chart to a region close to the 
line of $\beta$-stability. In contrast, most of our data are for
more neutron-deficient nuclei.
More neutron-deficient nuclei require longer evaporation chains, on 
average, and consequently higher excitation energies of the corresponding 
prefragments, which is in turn related to a stronger slowing-down to 
convert kinetic energy into excitation energy.

\begin{figure}
\begin{center}
\leavevmode
\epsfxsize=10cm
\epsffile{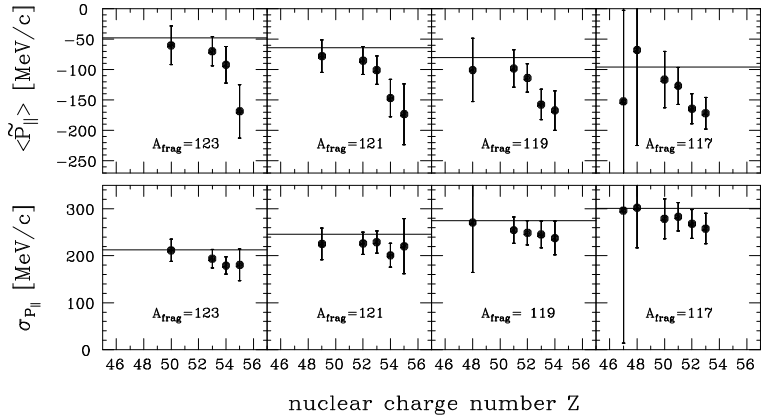}
\end{center}
\caption{Isobaric distributions (note that $\left< \tilde
P_{\Vert}\right>$ is not the real fragment momentum but its
mean velocity multiplied with a constant factor. For details
see text.) The full lines show the empirical
parameterizations by Morrissey~\protect\cite{morrissey}.}
\label{fig:momentum_isobar} 
\end{figure}

To illustrate this in more detail, we plot
in the upper part of Fig.~\ref{fig:momentum_isobar} the quantity
$\langle\tilde P_{\Vert}\rangle$ for four isobaric
distributions. The full line represents the prediction from the
systematics, $\langle\tilde P_{\Vert}\rangle=\Delta A\cdot 8MeV/c
$~\cite{morrissey}, which is a constant in isobaric chains. Clearly
visible is an increasing ``momentum transfer'' for the more neutron-deficient
isobars. This corroborates
our interpretation given above that those fragments are formed
via higher excitation energies of the corresponding prefragments and
subsequent emission of neutrons.
In contrast to that neutron-rich fragments
have to be formed with low excitation energies, because excitation of the
prefragment leads preferably to the emission of neutrons.
The same conclusion has been reached by Donzaud et al.~\cite{donzaud95} from a 
correlation of fragment longitudinal momenta with charged-particle
multiplicities.

The lower part of
Fig.~\ref{fig:momentum_isobar} shows the widths of the
momentum distributions for
the same four isobaric chains. In contrast to the "average momentum" they do
not exhibit such a significant dependence on the neutron-to-proton ratio.
It seems that the more neutron-deficient fragments have slightly
narrower momentum distributions. This is what one would expect from
the above discussion of the momentum transfer. The higher the
excitation energy, the more nucleons will be emitted during the deexcitation
phase. Assuming the Goldhaber model still holds for the temporarily formed
prefragments after the collision phase, the longitudinal
root-mean-square momentum per
abraded nucleon is approximately $P_{\Vert rms}\approx
\sqrt{1/5}P_{Fermi}=116$MeV/c (see Eq.~\ref{eqn:goldhaber}). This is
much larger, than the contribution from the evaporation of nucleons: the
emission of a nucleon with mass $m_N$ and kinetic energy of $E_{kin}\approx
2MeV$ results in 
$P_{\Vert rms}=\sqrt{1/3}\cdot\sqrt{2m_N\cdot E_{kin}} \approx 35MeV/c$.
Thus an increasing contribution of the evaporation phase to the observed 
mass-loss leads to narrower momentum distributions than expected from the 
Goldhaber model. This is corroborated by the observation that the momentum 
distribution of fragments that only have lost
protons~\cite{proton_removal,hanelt93} are satisfactorily 
described by the Goldhaber formalism as 
discussed above.
\section{Summary}
We have measured cross sections for projectile-like
fragments produced in the reaction $^{129}$Xe + $^{27}$Al at
an incident energy of 790A$\cdot$MeV. A comparison of the
isotopic distributions with those of neutron-rich fragments
observed in the fragmentation of $^{136}$Xe projectiles
reveals a dependence of the production yields on the
neutron-to-proton ratio of the incident projectile. This
so-called "memory effect" was so far only observed for
lighter projectiles. The very neutron-deficient tails of the
isotopic distributions indicate that production cross
sections for nuclei close to the proton dripline where
underestimated up to now. For most of the identified
fragments the longitudinal momentum distributions were
determined. They exhibit a behaviour in isobaric chains,
which is consistent with the assumption that
neutron-deficient fragments are formed from prefragments
with higher excitation energies and consequently larger
contributions from the subsequent evaporation cascade to the
observed mass-loss. This set of data now allows a detailed
comparison with microscopic descriptions, such as
intranuclear cascade models, which should allow a more
quantitative insight into the underlying reaction mechanism.
This will be presented in a forthcoming paper. 


The authors wish to thank K.-H.~Behr, A.~Br\"unle, and K.~Burkard for 
technical assistance in the preparation phase and during the experiment. 
Furthermore we would like to thank the accelerator crew of the GSI facility 
for providing a high quality $^{129}$Xe beam.

\begin{table}
\caption{\label{tbl:results} Measured cross sections and momentum
distributions (average momentum $<$P$>$ and rms width P$_{rms}$ measured
in the projectile frame) for the respective 
fragments formed in the reaction of 790 A$\cdot$MeV $^{129}$Xe + $^{27}$Al. 
For some isotopes only a cross section is given, as no
reliable fit of the momentum distribution was possible.
}
\begin{tabular}{ccdcr@{${}\pm{}$}rr@{${}\pm{}$}r}
El.& Z & A & cross section &  
\multicolumn{2}{c}{$<$P$>$} & 
\multicolumn{2}{c}{P$_{rms}$}\\
 & & &  [barn] & 
\multicolumn{2}{c}{[MeV/c]} &
\multicolumn{2}{c}{[MeV/c]} \\
\hline
 Cs & 55 & 129 & $( 8.53 \pm 0.80 )\cdot 10^{-4 }$& -177 & 90& 75 & 11 \\
 Cs & 55 & 128 & $( 2.29 \pm 0.19 )\cdot 10^{-3 }$& -161 & 41& 91 & 10 \\
 Cs & 55 & 127 & $( 4.18 \pm 0.30 )\cdot 10^{-3 }$& -168 & 86& 104 & 11 \\
 Cs & 55 & 126 & $( 3.83 \pm 0.36 )\cdot 10^{-3 }$& -196 & 43& 124 & 13 \\
 Cs & 55 & 125 & $( 3.62 \pm 0.34 )\cdot 10^{-3 }$& -207 & 76& 144 & 16 \\
 Cs & 55 & 124 & $( 2.04 \pm 0.39 )\cdot 10^{-3 }$& -274 & 49& 140 & 21 \\
 Cs & 55 & 123 & $( 1.46 \pm 0.28 )\cdot 10^{-3 }$& -298 & 77& 180 & 33 \\
 Cs & 55 & 122 & $( 6.77 \pm 0.74 )\cdot 10^{-4 }$& -282 & 63& 182 & 34 \\
 Cs & 55 & 121 & $( 2.39 \pm 0.28 )\cdot 10^{-4 }$& -302 & 87& 220 & 58 \\
 Cs & 55 & 120 & $( 1.19 \pm 1.20 )\cdot 10^{-4 }$\\
 Cs & 55 & 119 & $( 1.36 \pm 0.69 )\cdot 10^{-5 }$\\
 Cs & 55 & 118 & $( 6.06 \pm 3.69 )\cdot 10^{-6 }$\\
 Xe & 54 & 126 & $( 3.12 \pm 0.24 )\cdot 10^{-2 }$& -96 & 54& 112 & 11 \\
 Xe & 54 & 125 & $( 2.33 \pm 0.18 )\cdot 10^{-2 }$& -106 & 58& 144 & 14 \\
 Xe & 54 & 124 & $( 1.98 \pm 0.16 )\cdot 10^{-2 }$& -159 & 58& 161 & 16 \\
 Xe & 54 & 123 & $( 1.54 \pm 0.12 )\cdot 10^{-2 }$& -162 & 53& 179 & 18 \\
 Xe & 54 & 122 & $( 9.38 \pm 1.71 )\cdot 10^{-3 }$& -248 & 61& 185 & 21 \\
 Xe & 54 & 121 & $( 6.11 \pm 1.15 )\cdot 10^{-3 }$& -255 & 53& 201 & 25 \\
 Xe & 54 & 120 & $( 3.88 \pm 0.28 )\cdot 10^{-3 }$& -251 & 73& 232 & 31 \\
 Xe & 54 & 119 & $( 1.28 \pm 0.11 )\cdot 10^{-3 }$& -286 & 55& 237 & 35 \\
 Xe & 54 & 118 & $( 4.21 \pm 0.81 )\cdot 10^{-4 }$\\
 Xe & 54 & 117 & $( 7.17 \pm 1.74 )\cdot 10^{-5 }$\\
 Xe & 54 & 116 & $( 1.18 \pm 1.45 )\cdot 10^{-5 }$\\
 Xe & 54 & 115 & $( 3.08 \pm 2.75 )\cdot 10^{-6 }$\\
 I & 53 & 124 & $( 2.23 \pm 0.14 )\cdot 10^{-2 }$& -118 & 74& 184 & 18 \\
 I & 53 & 123 & $( 2.43 \pm 0.15 )\cdot 10^{-2 }$& -124 & 42& 193 & 19 \\
 I & 53 & 122 & $( 1.93 \pm 0.16 )\cdot 10^{-2 }$& -161 & 78& 215 & 22 \\
 I & 53 & 121 & $( 2.07 \pm 0.17 )\cdot 10^{-2 }$& -176 & 40& 229 & 23 \\
 I & 53 & 120 & $( 1.40 \pm 0.25 )\cdot 10^{-2 }$& -258 & 80& 227 & 28 \\
 I & 53 & 119 & $( 1.16 \pm 0.20 )\cdot 10^{-2 }$& -269 & 43& 245 & 28 \\
 I & 53 & 118 & $( 7.23 \pm 1.57 )\cdot 10^{-3 }$& -259 & 98& 277 & 42 \\
 I & 53 & 117 & $( 2.45 \pm 0.24 )\cdot 10^{-3 }$& -289 & 43& 257 & 32 \\
 I & 53 & 116 & $( 1.05 \pm 0.21 )\cdot 10^{-3 }$& -230 & 108& 220 & 54 \\
 I & 53 & 115 & $( 2.85 \pm 0.38 )\cdot 10^{-4 }$& -366 & 102& 306 & 95 \\
 I & 53 & 114 & $( 4.19 \pm 4.50 )\cdot 10^{-5 }$\\
 I & 53 & 113 & $( 7.20 \pm 8.80 )\cdot 10^{-6 }$\\
\end{tabular}
\end{table}
\addtocounter{table}{-1}
\begin{table}
\caption{(continued)}
\begin{tabular}{ccdcr@{${}\pm{}$}rr@{${}\pm{}$}r}
El.& Z & A & cross section &  
\multicolumn{2}{c}{$<$P$>$} & 
\multicolumn{2}{c}{P$_{rms}$}\\
 & & &  [barn] & 
\multicolumn{2}{c}{[MeV/c]} &
\multicolumn{2}{c}{[MeV/c]} \\
\hline
 Te & 52 & 128 & $( 1.65 \pm 0.49 )\cdot 10^{-5 }$& -98 & 55& 94 & 11 \\
 Te & 52 & 127 & $( 3.67 \pm 0.55 )\cdot 10^{-4 }$\\
 Te & 52 & 122 & $( 1.07 \pm 0.11 )\cdot 10^{-2 }$& -181 & 95& 190 & 23 \\
 Te & 52 & 121 & $( 1.40 \pm 0.09 )\cdot 10^{-2 }$& -148 & 40& 226 & 23 \\
 Te & 52 & 120 & $( 1.65 \pm 0.11 )\cdot 10^{-2 }$& -158 & 81& 251 & 27 \\
 Te & 52 & 119 & $( 1.65 \pm 0.13 )\cdot 10^{-2 }$& -195 & 39& 248 & 25 \\
 Te & 52 & 118 & $( 1.77 \pm 0.15 )\cdot 10^{-2 }$& -201 & 79& 273 & 29 \\
 Te & 52 & 117 & $( 1.39 \pm 0.24 )\cdot 10^{-2 }$& -276 & 41& 268 & 30 \\
 Te & 52 & 116 & $( 1.10 \pm 0.20 )\cdot 10^{-2 }$& -276 & 88& 288 & 42 \\
 Te & 52 & 115 & $( 6.01 \pm 0.50 )\cdot 10^{-3 }$& -292 & 42& 265 & 30 \\
 Te & 52 & 114 & $( 3.13 \pm 0.30 )\cdot 10^{-3 }$& -279 & 84& 274 & 48 \\
 Te & 52 & 113 & $( 6.07 \pm 0.84 )\cdot 10^{-4 }$& -380 & 79& 336 & 66 \\
 Te & 52 & 112 & $( 2.97 \pm 1.09 )\cdot 10^{-4 }$\\
 Te & 52 & 111 & $( 2.42 \pm 0.96 )\cdot 10^{-5 }$\\
 Te & 52 & 110 & $( 2.01 \pm 3.91 )\cdot 10^{-5 }$\\
 Sb & 51 & 127 & $( 2.46 \pm 2.25 )\cdot 10^{-7 }$\\
 Sb & 51 & 126 & $( 1.72 \pm 0.27 )\cdot 10^{-5 }$& -89 & 41& 151 & 17 \\
 Sb & 51 & 125 & $( 1.32 \pm 0.16 )\cdot 10^{-4 }$& -82 & 85& 175 & 18 \\
 Sb & 51 & 119 & $( 7.89 \pm 0.54 )\cdot 10^{-3 }$& -168 & 52& 254 & 28 \\
 Sb & 51 & 118 & $( 1.09 \pm 0.07 )\cdot 10^{-2 }$& -189 & 61& 297 & 32 \\
 Sb & 51 & 117 & $( 1.30 \pm 0.11 )\cdot 10^{-2 }$& -213 & 50& 282 & 30 \\
 Sb & 51 & 116 & $( 1.43 \pm 0.12 )\cdot 10^{-2 }$& -217 & 60& 304 & 32 \\
 Sb & 51 & 115 & $( 1.49 \pm 0.26 )\cdot 10^{-2 }$& -319 & 48& 296 & 34 \\
 Sb & 51 & 114 & $( 1.30 \pm 0.23 )\cdot 10^{-2 }$& -310 & 70& 316 & 40 \\
 Sb & 51 & 113 & $( 7.94 \pm 0.59 )\cdot 10^{-3 }$& -306 & 50& 321 & 36 \\
 Sb & 51 & 112 & $( 5.21 \pm 0.41 )\cdot 10^{-3 }$& -303 & 69& 333 & 49 \\
 Sb & 51 & 111 & $( 1.40 \pm 0.18 )\cdot 10^{-3 }$& -405 & 74& 370 & 66 \\
 Sb & 51 & 110 & $( 5.45 \pm 1.05 )\cdot 10^{-4 }$& -368 & 182& 388 & 195 \\
 Sb & 51 & 109 & $( 6.84 \pm 1.14 )\cdot 10^{-5 }$\\
 Sb & 51 & 108 & $( 1.02 \pm 0.83 )\cdot 10^{-5 }$\\
 Sn & 50 & 124 & $( 4.65 \pm 0.65 )\cdot 10^{-6 }$& -127 & 61& 188 & 33 \\
 Sn & 50 & 123 & $( 2.70 \pm 0.32 )\cdot 10^{-5 }$& -106 & 56& 211 & 23 \\
 Sn & 50 & 122 & $( 4.98 \pm 1.07 )\cdot 10^{-5 }$\\
 Sn & 50 & 117 & $( 4.44 \pm 0.34 )\cdot 10^{-3 }$& -196 & 77& 278 & 42 \\
 Sn & 50 & 116 & $( 5.79 \pm 0.40 )\cdot 10^{-3 }$& -174 & 44& 285 & 33 \\
 Sn & 50 & 115 & $( 8.73 \pm 0.77 )\cdot 10^{-3 }$& -244 & 70& 316 & 39 \\
 Sn & 50 & 114 & $( 1.16 \pm 0.09 )\cdot 10^{-2 }$& -215 & 44& 304 & 32 \\
 Sn & 50 & 113 & $( 1.20 \pm 0.23 )\cdot 10^{-2 }$& -300 & 69& 326 & 45 \\
 Sn & 50 & 112 & $( 1.32 \pm 0.23 )\cdot 10^{-2 }$& -307 & 53& 334 & 41 \\
 Sn & 50 & 111 & $( 1.14 \pm 0.09 )\cdot 10^{-2 }$& -310 & 65& 357 & 43 \\
\end{tabular}
\end{table}
\addtocounter{table}{-1}
\begin{table}
\caption{(continued)}
\begin{tabular}{ccdcr@{${}\pm{}$}rr@{${}\pm{}$}r}
El.& Z & A & cross section &  
\multicolumn{2}{c}{$<$P$>$} & 
\multicolumn{2}{c}{P$_{rms}$}\\
 & & &  [barn] & 
\multicolumn{2}{c}{[MeV/c]} &
\multicolumn{2}{c}{[MeV/c]} \\
\hline
 Sn & 50 & 110 & $( 8.22 \pm 0.53 )\cdot 10^{-3 }$& -314 & 53& 340 & 41 \\
 Sn & 50 & 109 & $( 2.88 \pm 0.36 )\cdot 10^{-3 }$& -429 & 63& 327 & 52 \\
 Sn & 50 & 108 & $( 1.27 \pm 0.17 )\cdot 10^{-3 }$& -432 & 72& 359 & 101 \\
 Sn & 50 & 107 & $( 2.40 \pm 0.22 )\cdot 10^{-4 }$& -365 & 107& 432 & 89 \\
 Sn & 50 & 106 & $( 3.80 \pm 0.51 )\cdot 10^{-5 }$\\
 Sn & 50 & 105 & $( 4.23 \pm 0.95 )\cdot 10^{-6 }$\\
 Sn & 50 & 104 & $( 5.77 \pm 3.61 )\cdot 10^{-7 }$\\
 Sn & 50 & 102 & $( 1.12 \pm 2.00 )\cdot 10^{-9 }$\\
 In & 49 & 122 & $( 1.50 \pm 0.29 )\cdot 10^{-6 }$\\
 In & 49 & 121 & $( 8.86 \pm 1.15 )\cdot 10^{-6 }$& -135 & 46& 225 & 33 \\
 In & 49 & 120 & $( 2.41 \pm 0.32 )\cdot 10^{-5 }$& -110 & 94& 257 & 36 \\
 In & 49 & 115 & $( 1.80 \pm 0.26 )\cdot 10^{-3 }$\\
 In & 49 & 114 & $( 3.53 \pm 0.25 )\cdot 10^{-3 }$\\
 In & 49 & 113 & $( 4.91 \pm 0.39 )\cdot 10^{-3 }$\\
 In & 49 & 112 & $( 8.12 \pm 0.66 )\cdot 10^{-3 }$\\
 In & 49 & 111 & $( 1.09 \pm 0.20 )\cdot 10^{-2 }$\\
 In & 49 & 110 & $( 1.03 \pm 0.18 )\cdot 10^{-2 }$\\
 In & 49 & 109 & $( 1.25 \pm 0.17 )\cdot 10^{-2 }$& -354 & 89& 374 & 52 \\
 In & 49 & 108 & $( 8.42 \pm 0.60 )\cdot 10^{-3 }$& -333 & 45& 367 & 42 \\
 In & 49 & 107 & $( 5.16 \pm 0.65 )\cdot 10^{-3 }$& -442 & 93& 386 & 64 \\
 In & 49 & 106 & $( 1.77 \pm 0.22 )\cdot 10^{-3 }$& -356 & 77& 412 & 77 \\
 In & 49 & 105 & $( 5.61 \pm 0.45 )\cdot 10^{-4 }$& -391 & 97& 425 & 69 \\
 In & 49 & 104 & $( 1.00 \pm 0.10 )\cdot 10^{-4 }$& -295 & 129& 478 & 143 \\
 In & 49 & 103 & $( 1.04 \pm 0.18 )\cdot 10^{-5 }$\\
 In & 49 & 102 & $( 0.43 \pm 1.92 )\cdot 10^{-5 }$\\
 Cd & 48 & 119 & $( 2.02 \pm 0.31 )\cdot 10^{-6 }$& -172 & 88& 270 & 106 \\
 Cd & 48 & 118 & $( 9.16 \pm 1.19 )\cdot 10^{-6 }$& -137 & 65& 265 & 44 \\
 Cd & 48 & 112 & $( 1.94 \pm 0.15 )\cdot 10^{-3 }$\\
 Cd & 48 & 111 & $( 2.96 \pm 0.22 )\cdot 10^{-3 }$\\
 Cd & 48 & 110 & $( 7.28 \pm 0.63 )\cdot 10^{-3 }$\\
 Cd & 48 & 109 & $( 9.57 \pm 2.45 )\cdot 10^{-3 }$\\
 Cd & 48 & 108 & $( 9.73 \pm 1.72 )\cdot 10^{-3 }$\\
 Cd & 48 & 107 & $( 1.10 \pm 0.21 )\cdot 10^{-2 }$& -370 & 144& 394 & 70 \\
 Cd & 48 & 106 & $( 9.17 \pm 0.68 )\cdot 10^{-3 }$& -353 & 43& 404 & 47 \\
 Cd & 48 & 105 & $( 9.18 \pm 1.03 )\cdot 10^{-3 }$& -492 & 102& 360 & 63 \\
 Cd & 48 & 104 & $( 3.22 \pm 0.38 )\cdot 10^{-3 }$& -407 & 61& 413 & 63 \\
 Cd & 48 & 103 & $( 1.04 \pm 0.10 )\cdot 10^{-3 }$& -418 & 100& 426 & 63 \\
 Cd & 48 & 102 & $( 2.17 \pm 0.17 )\cdot 10^{-4 }$& -293 & 66& 380 & 67 \\
 Cd & 48 & 101 & $( 2.56 \pm 0.33 )\cdot 10^{-5 }$& -402 & 55& 467 & 48 \\
 Cd & 48 & 100 & $( 2.75 \pm 0.92 )\cdot 10^{-6 }$\\
\end{tabular}
\end{table}
\addtocounter{table}{-1}
\begin{table}
\caption{(continued)}
\begin{tabular}{ccdcr@{${}\pm{}$}rr@{${}\pm{}$}r}
El.& Z & A & cross section &  
\multicolumn{2}{c}{$<$P$>$} & 
\multicolumn{2}{c}{P$_{rms}$}\\
 & & &  [barn] & 
\multicolumn{2}{c}{[MeV/c]} &
\multicolumn{2}{c}{[MeV/c]} \\
\hline
 Ag & 47 & 117 & $( 1.62 \pm 0.33 )\cdot 10^{-6 }$& -256 & 252& 296 & 282 \\
 Ag & 47 & 116 & $( 5.30 \pm 0.74 )\cdot 10^{-6 }$& -181 & 77& 256 & 93 \\
 Ag & 47 & 115 & $( 1.87 \pm 0.27 )\cdot 10^{-5 }$& -112 & 164& 300 & 87 \\
 Ag & 47 & 110 & $( 8.55 \pm 1.00 )\cdot 10^{-4 }$\\
 Ag & 47 & 109 & $( 2.23 \pm 0.16 )\cdot 10^{-3 }$\\
 Ag & 47 & 108 & $( 3.76 \pm 0.34 )\cdot 10^{-3 }$\\
 Ag & 47 & 107 & $( 5.77 \pm 0.48 )\cdot 10^{-3 }$\\
 Ag & 47 & 106 & $( 8.24 \pm 1.45 )\cdot 10^{-3 }$\\
 Ag & 47 & 105 & $( 9.33 \pm 1.65 )\cdot 10^{-3 }$\\
 Ag & 47 & 104 & $( 9.06 \pm 0.64 )\cdot 10^{-3 }$& -378 & 47& 400 & 48 \\
 Ag & 47 & 103 & $( 7.62 \pm 0.70 )\cdot 10^{-3 }$& -353 & 115& 447 & 80 \\
 Ag & 47 & 102 & $( 4.13 \pm 0.48 )\cdot 10^{-3 }$& -401 & 60& 444 & 68 \\
 Ag & 47 & 101 & $( 1.86 \pm 0.24 )\cdot 10^{-3 }$& -429 & 118& 447 & 66 \\
 Ag & 47 & 100 & $( 4.43 \pm 0.36 )\cdot 10^{-4 }$& -337 & 66& 484 & 83 \\
 Ag & 47 & 99 & $( 5.46 \pm 0.51 )\cdot 10^{-5 }$& -427 & 201& 480 & 124 \\
 Ag & 47 & 98 & $( 6.14 \pm 1.17 )\cdot 10^{-6 }$\\
 Pd & 46 & 114 & $( 1.11 \pm 0.20 )\cdot 10^{-6 }$\\
 Pd & 46 & 113 & $( 4.64 \pm 0.71 )\cdot 10^{-6 }$& -208 & 155& 337 & 160 \\
 Pd & 46 & 112 & $( 8.87 \pm 3.64 )\cdot 10^{-6 }$\\
 Pd & 46 & 107 & $( 1.09 \pm 0.09 )\cdot 10^{-3 }$\\
 Pd & 46 & 106 & $( 2.27 \pm 0.26 )\cdot 10^{-3 }$\\
 Pd & 46 & 105 & $( 4.04 \pm 0.34 )\cdot 10^{-3 }$\\
 Pd & 46 & 104 & $( 6.77 \pm 1.23 )\cdot 10^{-3 }$\\
 Pd & 46 & 103 & $( 8.13 \pm 1.43 )\cdot 10^{-3 }$\\
 Pd & 46 & 102 & $( 9.09 \pm 0.63 )\cdot 10^{-3 }$& -370 & 68& 431 & 61 \\
 Pd & 46 & 101 & $( 8.30 \pm 0.59 )\cdot 10^{-3 }$& -356 & 83& 445 & 66 \\
 Pd & 46 & 100 & $( 5.73 \pm 0.67 )\cdot 10^{-3 }$& -438 & 59& 432 & 62 \\
 Pd & 46 & 99 & $( 2.13 \pm 0.28 )\cdot 10^{-3 }$& -331 & 171& 441 & 137 \\
 Pd & 46 & 98 & $( 7.14 \pm 0.55 )\cdot 10^{-4 }$& -369 & 58& 495 & 76 \\
 Pd & 46 & 97 & $( 1.02 \pm 0.09 )\cdot 10^{-4 }$& -473 & 167& 460 & 99 \\
 Pd & 46 & 96 & $( 1.11 \pm 0.14 )\cdot 10^{-5 }$\\
 Rh & 45 & 112 & $( 8.78 \pm 21.45 )\cdot 10^{-7 }$\\
 Rh & 45 & 111 & $( 2.14 \pm 0.32 )\cdot 10^{-6 }$\\
 Rh & 45 & 110 & $( 5.23 \pm 0.82 )\cdot 10^{-6 }$\\
 Rh & 45 & 105 & $( 5.74 \pm 0.55 )\cdot 10^{-4 }$\\
 Rh & 45 & 104 & $( 1.22 \pm 0.10 )\cdot 10^{-3 }$\\
 Rh & 45 & 103 & $( 2.85 \pm 0.25 )\cdot 10^{-3 }$\\
 Rh & 45 & 102 & $( 3.47 \pm 0.33 )\cdot 10^{-3 }$\\
 Rh & 45 & 101 & $( 6.63 \pm 1.17 )\cdot 10^{-3 }$\\
 Rh & 45 & 100 & $( 8.15 \pm 0.64 )\cdot 10^{-3 }$& -395 & 105& 424 & 68 \\
\end{tabular}
\addtocounter{table}{-1}
\end{table}
\begin{table}
\caption{(continued)}
\begin{tabular}{ccdcr@{${}\pm{}$}rr@{${}\pm{}$}r}
El.& Z & A & cross section &  
\multicolumn{2}{c}{$<$P$>$} & 
\multicolumn{2}{c}{P$_{rms}$}\\
 & & &  [barn] & 
\multicolumn{2}{c}{[MeV/c]} &
\multicolumn{2}{c}{[MeV/c]} \\
\hline
 Rh & 45 & 99 & $( 7.89 \pm 0.53 )\cdot 10^{-3 }$& -357 & 62& 454 & 65 \\
 Rh & 45 & 98 & $( 6.38 \pm 0.76 )\cdot 10^{-3 }$& -449 & 76& 436 & 68 \\
 Rh & 45 & 97 & $( 3.02 \pm 0.37 )\cdot 10^{-3 }$& -407 & 150& 483 & 135 \\
 Rh & 45 & 96 & $( 1.06 \pm 0.07 )\cdot 10^{-3 }$& -405 & 56& 480 & 69 \\
 Rh & 45 & 95 & $( 1.50 \pm 0.14 )\cdot 10^{-4 }$& -465 & 198& 490 & 112 \\
 Rh & 45 & 94 & $( 2.37 \pm 0.26 )\cdot 10^{-5 }$& -377 & 161& 506 & 204 \\
 Ru & 44 & 103 & $( 4.46 \pm 0.69 )\cdot 10^{-4 }$\\
 Ru & 44 & 102 & $( 6.89 \pm 0.61 )\cdot 10^{-4 }$\\
 Ru & 44 & 101 & $( 1.67 \pm 0.35 )\cdot 10^{-3 }$\\
 Ru & 44 & 100 & $( 2.80 \pm 0.25 )\cdot 10^{-3 }$\\
 Ru & 44 & 99 & $( 5.28 \pm 0.94 )\cdot 10^{-3 }$\\
 Ru & 44 & 98 & $( 6.06 \pm 1.17 )\cdot 10^{-3 }$& -521 & 127& 400 & 81 \\
 Ru & 44 & 97 & $( 7.41 \pm 0.49 )\cdot 10^{-3 }$& -358 & 51& 447 & 63 \\
 Ru & 44 & 96 & $( 7.27 \pm 0.90 )\cdot 10^{-3 }$& -453 & 109& 455 & 82 \\
 Ru & 44 & 95 & $( 3.03 \pm 0.73 )\cdot 10^{-3 }$& -402 & 93& 459 & 99 \\
 Ru & 44 & 94 & $( 1.47 \pm 0.10 )\cdot 10^{-3 }$& -404 & 63& 499 & 70 \\
 Ru & 44 & 93 & $( 2.24 \pm 0.24 )\cdot 10^{-4 }$& -525 & 197& 491 & 100 \\
 Ru & 44 & 92 & $( 3.14 \pm 0.30 )\cdot 10^{-5 }$& -385 & 151& 559 & 218 \\
 Tc & 43 & 100 & $( 4.23 \pm 0.45 )\cdot 10^{-4 }$\\
 Tc & 43 & 99 & $( 9.91 \pm 1.27 )\cdot 10^{-4 }$\\
 Tc & 43 & 98 & $( 2.01 \pm 0.19 )\cdot 10^{-3 }$\\
 Tc & 43 & 97 & $( 4.39 \pm 0.79 )\cdot 10^{-3 }$\\
 Tc & 43 & 96 & $( 5.00 \pm 0.91 )\cdot 10^{-3 }$\\
 Tc & 43 & 95 & $( 7.24 \pm 0.48 )\cdot 10^{-3 }$& -365 & 53& 437 & 64 \\
 Tc & 43 & 94 & $( 5.50 \pm 0.51 )\cdot 10^{-3 }$& -559 & 120& 429 & 83 \\
 Tc & 43 & 93 & $( 4.81 \pm 0.57 )\cdot 10^{-3 }$& -394 & 76& 485 & 92 \\
 Tc & 43 & 92 & $( 1.78 \pm 0.12 )\cdot 10^{-3 }$& -451 & 70& 478 & 66 \\
 Tc & 43 & 91 & $( 3.50 \pm 0.37 )\cdot 10^{-4 }$& -316 & 140& 493 & 125 \\
 Tc & 43 & 90 & $( 3.77 \pm 0.35 )\cdot 10^{-5 }$& -403 & 133& 531 & 178 \\
 Mo & 42 & 98 & $( 2.41 \pm 0.31 )\cdot 10^{-4 }$\\
 Mo & 42 & 97 & $( 5.20 \pm 0.57 )\cdot 10^{-4 }$\\
 Mo & 42 & 96 & $( 1.43 \pm 0.14 )\cdot 10^{-3 }$\\
 Mo & 42 & 95 & $( 2.47 \pm 0.48 )\cdot 10^{-3 }$\\
 Mo & 42 & 94 & $( 4.99 \pm 0.90 )\cdot 10^{-3 }$\\
 Mo & 42 & 93 & $( 6.10 \pm 0.44 )\cdot 10^{-3 }$& -411 & 78& 464 & 81 \\
 Mo & 42 & 92 & $( 6.52 \pm 0.56 )\cdot 10^{-3 }$& -345 & 137& 479 & 103 \\
 Mo & 42 & 91 & $( 5.32 \pm 0.63 )\cdot 10^{-3 }$& -404 & 80& 473 & 111 \\
 Mo & 42 & 90 & $( 1.93 \pm 0.15 )\cdot 10^{-3 }$& -479 & 89& 497 & 73 \\
 Mo & 42 & 89 & $( 4.37 \pm 0.34 )\cdot 10^{-4 }$& -328 & 118& 529 & 130 \\
 Mo & 42 & 88 & $( 7.30 \pm 0.59 )\cdot 10^{-5 }$& -422 & 132& 546 & 165 \\
\end{tabular}
\addtocounter{table}{-1}
\end{table}
\begin{table}
\caption{(continued)}
\begin{tabular}{ccdcr@{${}\pm{}$}rr@{${}\pm{}$}r}
El.& Z & A & cross section &  
\multicolumn{2}{c}{$<$P$>$} & 
\multicolumn{2}{c}{P$_{rms}$}\\
 & & &  [barn] & 
\multicolumn{2}{c}{[MeV/c]} &
\multicolumn{2}{c}{[MeV/c]} \\
\hline
 Nb & 41 & 96 & $( 1.67 \pm 0.72 )\cdot 10^{-4 }$\\
 Nb & 41 & 95 & $( 2.75 \pm 0.32 )\cdot 10^{-4 }$\\
 Nb & 41 & 94 & $( 9.47 \pm 1.03 )\cdot 10^{-4 }$\\
 Nb & 41 & 93 & $( 1.80 \pm 0.18 )\cdot 10^{-3 }$\\
 Nb & 41 & 92 & $( 3.83 \pm 0.69 )\cdot 10^{-3 }$\\
 Nb & 41 & 91 & $( 5.56 \pm 0.42 )\cdot 10^{-3 }$& -427 & 124& 476 & 110 \\
 Nb & 41 & 90 & $( 6.14 \pm 0.48 )\cdot 10^{-3 }$& -387 & 112& 523 & 139 \\
 Nb & 41 & 89 & $( 5.56 \pm 0.65 )\cdot 10^{-3 }$& -458 & 70& 470 & 92 \\
 Nb & 41 & 88 & $( 1.75 \pm 0.27 )\cdot 10^{-3 }$& -502 & 113& 507 & 79 \\
 Nb & 41 & 87 & $( 5.70 \pm 0.44 )\cdot 10^{-4 }$& -389 & 100& 542 & 129 \\
 Nb & 41 & 86 & $( 1.01 \pm 0.09 )\cdot 10^{-4 }$& -415 & 155& 565 & 166 \\
 Zr & 40 & 93 & $( 1.61 \pm 0.24 )\cdot 10^{-4 }$\\
 Zr & 40 & 92 & $( 4.58 \pm 0.76 )\cdot 10^{-4 }$\\
 Zr & 40 & 91 & $( 1.16 \pm 0.13 )\cdot 10^{-3 }$\\
 Zr & 40 & 90 & $( 3.49 \pm 0.65 )\cdot 10^{-3 }$\\
 Zr & 40 & 89 & $( 4.45 \pm 0.45 )\cdot 10^{-3 }$& -540 & 141& 411 & 102 \\
 Zr & 40 & 88 & $( 5.88 \pm 0.43 )\cdot 10^{-3 }$& -330 & 71& 497 & 111 \\
 Zr & 40 & 87 & $( 5.81 \pm 0.71 )\cdot 10^{-3 }$& -463 & 105& 501 & 115 \\
 Zr & 40 & 86 & $( 2.24 \pm 0.23 )\cdot 10^{-3 }$& -561 & 127& 491 & 80 \\
 Zr & 40 & 85 & $( 7.14 \pm 0.53 )\cdot 10^{-4 }$& -419 & 71& 505 & 103 \\
\\
\end{tabular}
\end{table}

\end{document}